\documentclass[3p]{elsarticle}

\usepackage{lineno,hyperref}
\usepackage{xspace}
\usepackage{enumerate}
\usepackage{graphicx} 
\usepackage{booktabs}
\usepackage{threeparttable}
\usepackage{multicol}
\usepackage{multirow}
\usepackage{lipsum}
\usepackage{setspace}
\usepackage{pdfpages}
\usepackage{algorithm}  
\usepackage{algpseudocode}  
\usepackage{float}
\usepackage{subfigure} 
\usepackage{array}
\usepackage{longtable}
\usepackage{amsmath}
\usepackage{hyperref}
\usepackage{supertabular}

\hypersetup{
    colorlinks=true,
    linkcolor=blue,
    filecolor=blue,      
    urlcolor=blue,
    citecolor=cyan,
}

\modulolinenumbers[5]

\journal{Journal of \LaTeX\ Templates}

\begin{document}

\begin{frontmatter}

\title{Learning Robust Heterogeneous Signal Features from Parallel Neural Network for Audio Sentiment Analysis}

\author[1,3]{{Feiyang Chen}\corref{mycorrespondingauthor}}
\cortext[mycorrespondingauthor]{Feiyang Chen and Ziqian Luo's work as Research Intern at State Key Laboratory of Intelligent Technology and Systems, Tsinghua University, advised by Prof. Hua Xu}
\ead{chenfeiyang999@gmail.com}
\author[2,3]{Ziqian Luo}
\ead{luoziqian98@gmail.com}
\address[1]{Department of Computer Science and Technology, 
  Beijing Forestry University, Beijing 100083, China}
\address[2]{Department of Communication Engineering, 
  Beijing University of Posts and Telecommunications, 
  Beijing 100876, China}
\address[3]{State Key Laboratory of Intelligent Technology and Systems,
 Department of Computer Science and Technology,\\ 
 Tsinghua University, Beijing 100084, China}

\begin{abstract}
Audio Sentiment Analysis is a popular research area which extends the conventional text-based sentiment analysis to depend on the effectiveness of acoustic features extracted from speech. However, current progress on audio sentiment analysis mainly focuses on extracting homogeneous acoustic features or doesn't fuse heterogeneous features effectively. In this paper, we propose an utterance-based deep neural network model, which has a parallel combination of Convolutional Neural Network (CNN) and Long Short-Term Memory (LSTM) based network, to obtain representative features termed Audio Sentiment Vector (ASV), that can maximally reflect sentiment information in an audio. Specifically, our model is trained by utterance-level labels and ASV can be extracted and fused creatively from two branches. In the CNN model branch, spectrum graphs produced by signals are fed as inputs while in the LSTM model branch, inputs include spectral features and cepstrum coefficient extracted from dependent utterances in an audio. Besides, Bidirectional Long Short-Term Memory (BiLSTM) with attention mechanism is used for feature fusion. Extensive experiments have been conducted to show our model can recognize audio sentiment precisely and quickly, and demonstrate our ASV are better than traditional acoustic features or vectors extracted from other deep learning models. Furthermore, experimental results indicate that the proposed model outperforms the state-of-the-art approach by 9.33\% on Multimodal Opinion-level Sentiment Intensity dataset (MOSI) dataset.
\end{abstract}

\begin{keyword}
\texttt{Audio Sentiment Analysis}\sep Feature Fusion \sep Signal Processing
\MSC[2018] 00-01\sep  99-00
\end{keyword}

\end{frontmatter}


\section{Introduction}

Sentiment Analysis is a well-studied research area in Natural Language Processing (NLP) \cite{pang2008opinion}, which is the computational study of peoples' opinions, sentiments, appraisals, and attitudes towards entities such as products, services, organizations and so on \cite{liu2015sentiment,tao2012minimum}. Traditional sentiment analysis methods are mostly based on texts \cite{de1982texture,passalis2017neural}, with the rapid development of communication technology, abundance of smartphones and the rapid rise of social media, large amounts of data are uploaded by web users in the form of audios or videos \cite{reihanian2018overlapping,lin2017dynamic}, rather than texts \cite{poria2017review}. Interestingly, a recent study shows that voice-only as modality seems best for humans’ empathetic accuracy as compared to video-only or audiovisual communication \cite{kraus2017voice}. In fact, audio sentiment analysis is a difficult task due to the complexity of audio signal. It is generally known that speech is the most convenient and natural medium for human communication, not only carries the implicit semantic information, but also contains rich affective information \cite{zhang2018speech}. Therefore, audio sentiment analysis, which aims to correctly analyze the sentiment of the speaker from speech signals, has drawn a great deal of attention of researchers.

In recent years, there are three main methods for audio sentiment analysis. Firstly, utilizes Automatic Speech Recognition (ASR) technology to convert speech into texts, following by conventional text-based sentiment analysis systems \cite{ezzat2012sentiment}. Secondly, adopts a generative model operating directly on the raw audio waveform \cite{van2016wavenet}. Thirdly, focuses on extracting signal features from the raw audio files \cite{bertin2011large}, which well captures the tonal content of a music, and has been proved to be more effective than original audio spectrums descriptors such as Mel-Frequency Cepstrum Coefficients(MFCC).

However, for converting speech into texts, by recognizing each word said by the person in an audio, change them into the word embedding and use some techniques in NLP, like Term Frequency–Inverse Document Frequency (TF-IDF) and Bag of Words (BOW) model \cite{passalis2017neural}. The result is not always accurate, because sentiment detection accuracy depends on being able to reliably detect a very focused vocabulary in the spoken comments \cite{kaushik2015automatic}. Furthermore, when the voice is transferred to the text, some sentiment-related signal characteristics are also lost, resulting in a decrease in the accuracy of the sentiment classification. As for extracting from the raw audio files through human works and then being put into the Support Vector Machine(SVM) classifier for classification, those methods require lots of human work and are heavily dependent on language types. 

Luckily, along with the success of deep learning in many other application domains, deep learning is also popularly used in audio sentiment analysis in recent years \cite{hong2018classification,mandanas2018m,mariel2018sentiment}. More recently, \cite{trigeorgis2016adieu} directly use the raw audio samples to train a Convolutional Recurrent Neural Network (CRNN) to predict continuous arousal /valence space. \cite{mirsamadi2017automatic} study the use of deep learning to automatically discover emotionally relevant features from speech. They propose a novel strategy for feature pooling over time which uses local attention in order to focus on specific regions of a speech signal that are more emotionally salient. \cite{neumann2017attentive} use an attentive convolutional neural network with multi-view learning objective function and achieved state-of-the-art results on the improvised speech data of \href{https://sail.usc.edu/iemocap/}{IEMOCAP} \cite{busso2008iemocap}. \cite{wang2017learning} propose to use Deep Neural Networks (DNN) to encode each utterance into a fixed-length vector by pooling the activations of the last hidden layer over time. The feature encoding process is designed to be jointly trained with the utterance-level classifier for better classification. \cite{chen20183} propose a 3-D attention-based convolutional recurrent neural networks to learn discriminative features for speech emotion recognition, where the Mel-spectrogram with deltas and delta-deltas are creatively used as input. But most of the previous methods still either considered only one single audio feature \cite{chen20183} or high-dimensional vectors \cite{lee2017scls,kim2014sentiment} from one homogeneous feature \cite{poria2017context}, and did not effectively extract and fuse audio features. 

We believe the information extracted from a single utterance must have dependency on its context. For example, a flash of loud expression may not indicate a person has a strong emotion since it maybe just caused by a cough while continuous loud one is far more likely to indicate the speaker has a strong emotion.

In this paper, based on a large number of experiments, we extract the features of each utterance in an audio through the \href{https://github.com/librosa/librosa}{Librosa} toolkit, and obtain four most effective features representing sentiment information, merge them by adopting a BiLSTM with attention mechanism. Moreover, we design a novel model called Audio Feature Fusion-Attention based CNN and RNN (AFF-ACRNN) for audio sentiment analysis. Spectrum graphs and selected traditional acoustic features are fed as input in two separate branches, we can obtain a new fusion of audio feature vector before the softmax layer, which we call the Audio Sentiment Vector (ASV). Finally, the output of the softmax layer is the class of sentiment. 

Major contributions of the paper are that:
\begin{itemize}
\item We propose an effective AFF-ACRNN model for audio sentiment analysis, through combining multiple traditional acoustic features and spectrum graphs to learn more comprehensive sentiment information in audio.
\item Our model is language insensitive and pay more attention to acoustic features of the original audio rather than words recognized from the audio.
\item Experimental results indicate that the proposed method outperforms the state-of-the-art methods \cite{poria2017context} on Multimodal Corpus of Sentiment Intensity dataset(\href{https://github.com/A2Zadeh/CMU-MultimodalSDK}{MOSI}) and Multimodal Opinion Utterances Dataset(\href{http://web.eecs.umich.edu/~mihalcea/downloads.html}{MOUD}).

\end{itemize}

The rest of the paper is organized as follows. In the following section, we will review related work. In Section 3, we will exhibit more details of our methodology. In Section 4, experiments and results are presented, and conclusion follows in Section 5.

\section{Related Work}

Current state-of-the-art methods for audio sentiment analysis are mostly based on deep neural network. In this section, we briefly present the advances on audio sentiment analysis task by utilizing deep learning, and then we give a summary on the progress of extracting the audio feature representation.
\subsection{Long Short-Term Memory (LSTM)}
 It has been demonstrated that LSTM \cite{hochreiter1997long} are well-suited to make predictions based on time series data, by utilizing a cell to remember values over arbitrary time intervals and the three gates(input gate $i$, output gate $o$, forget gate $f$) to regulate the flow of information into and out of the cell, which can be described  as follows: 

\begin{equation}
{{\rm{f}}_t} = \sigma ({W_f} \cdot [{h_{t - 1}},{x_t}] + {b_f}) 
\end{equation}

\begin{equation}
{i_t} = \sigma ({W_i} \cdot [{h_{t - 1}},{x_t}] + {b_i})
\end{equation}

\begin{equation}
{o_t} = \sigma ({W_o} \cdot [{h_{t - 1}},{x_t}] + {b_o})
\end{equation}
where ${h_t} = {o_t} * \tanh ({C_t})$ is the output of the last cell and $x_t$ is the input of current cell. Besides, the current cell state $C_t$ can be updated by the following formula:

\begin{equation}
\mathop {{C_t}}\limits^ \sim   = \tanh ({W_c} \cdot [{h_{t - 1}},{x_t}] + {b_c})
\end{equation}

\begin{equation}
{C_t} = {f_t} * {C_{t - 1}} + {i_t} * \mathop {{C_t}}\limits^ \sim  
\end{equation}
where $C_{t-1}$ stands for the previous cell state.

One of the most effective variant of LSTM is the bidirectional LSTM. Each input sequence will be fed into both the forward and backward LSTM layers and thus a hidden layer receives an input by joining forward and backward LSTM layers.

\subsection{Convolutional Neural Network (CNN)}
CNN \cite{lecun1990handwritten} are well-known for extracting features from an image by using convolutional kernels and pooling layers to emulates the response of an individual to visual stimuli. Moreover, CNN has been successfully used not only for computer vision, but also for speech \cite{sainath2015deep}. For speech recognition, CNN is proved to be robust against noise compared to other DL models \cite{palaz2015analysis}. 
\subsection{Audio Feature Representation and Extraction}
Researchers have found pitch and energy related features playing a key role in affect recognition \cite{poria2017context}. Other features that have been used by some researchers for feature extraction include formants, MFCC, root-mean-square energy, spectral centroid and tonal centroid features. MFCC is the most recognized feature among the four and the mapping between the real frequency scale (Hz) and the perceived frequency scales (mels) is approximately linear below 1 KHz and logarithmic at higher frequency, and such an approximation is usually adopted in speech recognition. There relationship is modeled as the formula suggested below:

\begin{equation}
F_{mel}=2595log_{10}(1+\frac{F_{Hz}}{{700}})
\end{equation}

During the speech production, there are several utterances and for each utterance, the audio signal can be divided into several segments. Global features are calculated by measuring several statistics, e.g., average, mean, deviation of the local features. Global features are the most commonly used features in the literature. They are fast to compute and, as they are fewer in number compared to local features, the overall speed of computation is enhanced \cite{el2011survey}. However, there are some drawbacks of calculating global features, as some of them are only useful to detect affect of high arousal, e.g., anger and disgust. For lower arousal, global features are not effective, e.g. global features are less prominent to distinguish between anger and joy. Global features also lack temporal information and dependence between two segments in an utterance. In a recent study \cite{cummins2017image}, a new acoustic feature representation, denoted as deep spectrum features, derived from feeding spectrum graphs through a very deep image classification CNN and forming a feature vector from the activation of the last fully connected layer. Librosa \cite{mcfee2015librosa} is an open-source python package for music and audio analysis which is able to extract all the key features as elaborated above.

\section{Methodology}

In this section, we describe the proposed AFF-ACRNN model for audio sentiment analysis in details. We firstly introduce an overview of the whole neural network architecture. After that, two separate branches of AFF-ACRNN will be explained in details. Finally, we talk about the fusion mechanism used in our model.

\begin{figure}[htbp] 
\centering
\includegraphics[width=0.3\textwidth]{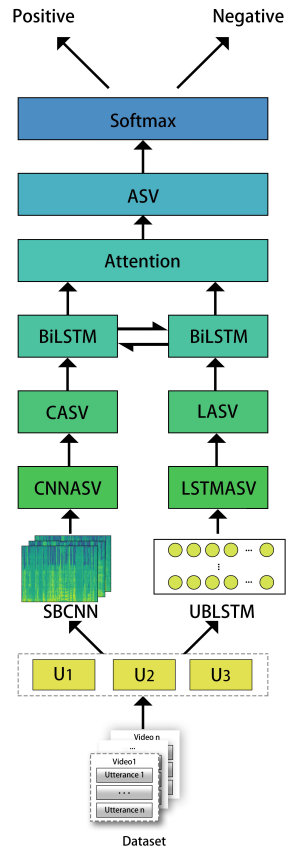} 
\caption{Overview of AFF-ACRNN Model} 
\label{Fig.main1} 
\end{figure}
\subsection{Model---AFF-ACRNN}

We concentrate on a model that has two parallel branches, the Utterance-Based BiLSTM Branch (UB-BiLSTM) and the Spectrum-Based CNN Branch (SBCNN), whose core mechanisms are based on LSTM and CNN. One branch of proposed model uses the BiLSTM to extract temporal information between adjacent utterances, another branch uses the renowned CNN based network to extract features from spectrum graph that sequence model cannot achieve. Furthermore, audio feature vector of each piece of utterance is the input of the proposed neural network that based on Audio Feature Fusion (AFF), we can obtain a new fusion audio feature vector before the softmax layer, which we call the Audio Sentiment Vector (ASV). Finally, the output of the softmax layer produces our final sentiment classification results, as shown in Figure 1.

\begin{figure}[htbp] 
\centering
\includegraphics[width=0.5\textwidth]{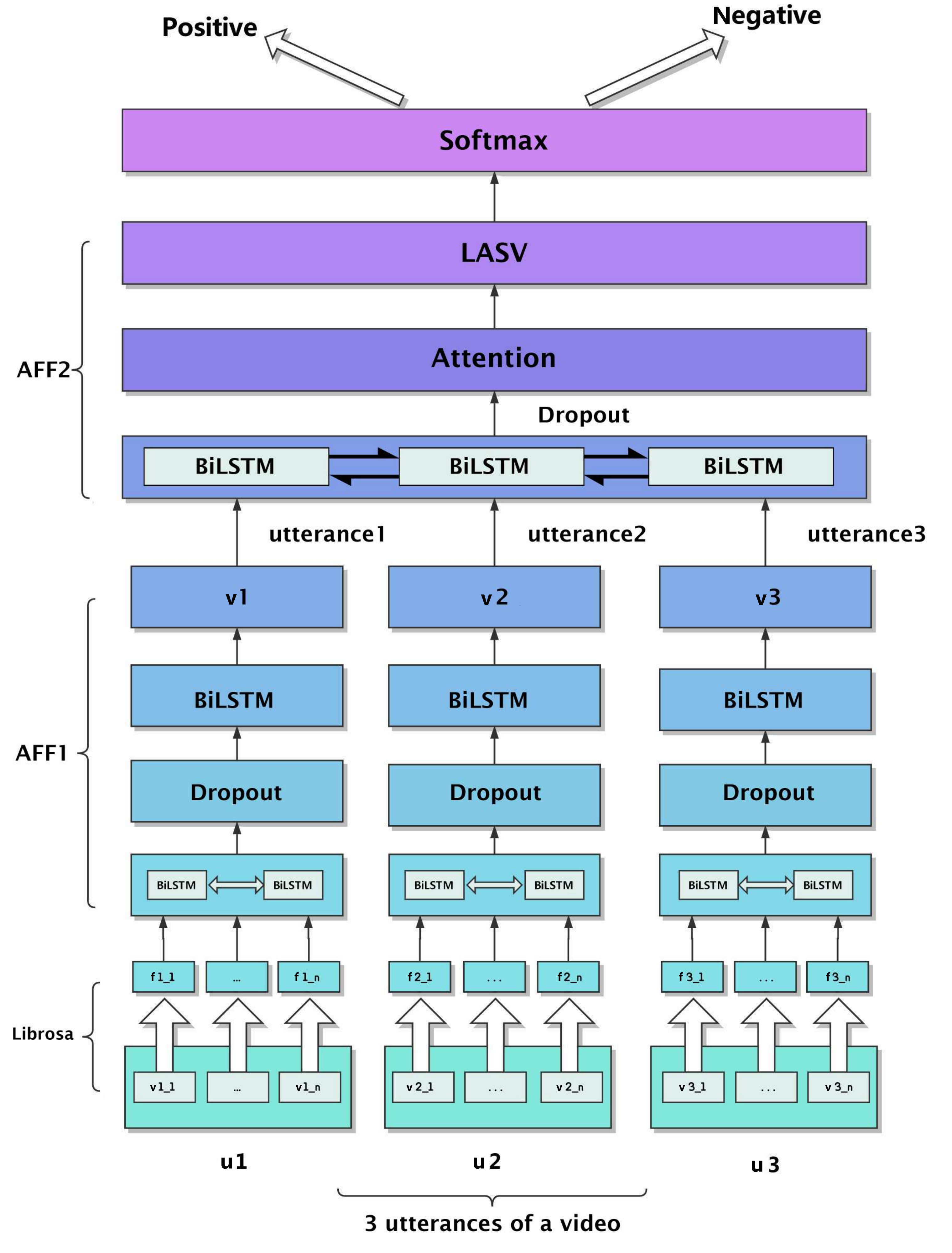} 
\caption{Overview of Our UB-BiLSTM Model} 
\label{Fig.main2} 
\end{figure}

\subsection{Audio Sentiment Vector (ASV) from Audio Feature Fusion (AFF)}
\subsubsection{LSTM Layers}

The hidden layers of LSTM have self-recurrent weights. These enable the cell in the memory block to retain previous information \cite{bae2016acoustic}. Firstly, we separate the different videos and take three continuous utterances (e.g. $u_1, u_2, u_3$) in one video at a time. Among them, for each utterance (e.g. $u_1$), we extract its internal acoustic features through the librosa toolkit, say $f_{1_1}, f_{1_2}...f_{1_n}$, and then trained by two layers of BiLSTM in AFF1 to obtain the extracted features from the traditional acoustic feature. Therefore, three utterances are corresponding to three more efficient and representative vectors $v_1, v_2, v_3$, as the inputs to BiLSTM in AFF2. AFF2 effectively combines the contextual information between adjacent utterances, and then subtly acquires the utterance that has the greatest impact on the final sentiment classification through the attention mechanism. Finally, after the dropout layer, a more representative vector, named as LASV is extracted by our LSTM framework before the softmax layer, as shown in Figure 2. The process is described in LSTM branch procedure in Algorithm 1.

\begin{algorithm}[ht]  
  \caption{Related Procedure}  
  \begin{algorithmic}[1]  
  \Procedure{LSTM branch}{}
    \For{i:[0,n]}  
      \State $f_i=getAudioFeature(u_i)$  
      \State $ASV_i=getASV(f_i)$   
    \EndFor 
    \For{i:[0,M]}  //M is the number of videos
      \State $input_i=GetTopUtter(v_i)$  
      \State $u_{f_i}=getUtterFeature(input_i)$   
    \EndFor  
\State $shuffle(v)$
\EndProcedure

  \Procedure{CNN Branch}{}
	\For{i:[0,n]}  
      \State $\textit{$x_i$} \gets \textit{get SpectrogramImage($u_i$)}$ 
      \State $\textit{$c_i$} \gets 	\textit{CNNModel($x_i$)}$   
      \State $\textit{$l_i$} \gets \textit{BiLSTM($c_i$)}$   
    \EndFor  
  \EndProcedure
  
  \Procedure{Find corresponding Label}{}
	\For{i:[0:2199]}  
      \State $rename(u_i)$    \qquad $//$ for better order in sorting
      \State $NameAndLabel=createIndex(u_i)$    
      \State $//$ A dictionary [utterance Name: Label]  
    \EndFor  
    \State $Label_x=NameAndLabel(u_x)$
  \EndProcedure
  \end{algorithmic}  
\end{algorithm}

\subsubsection{CNN Layers}

\begin{figure}[htbp] 
\centering
\includegraphics[width=0.26\textwidth]{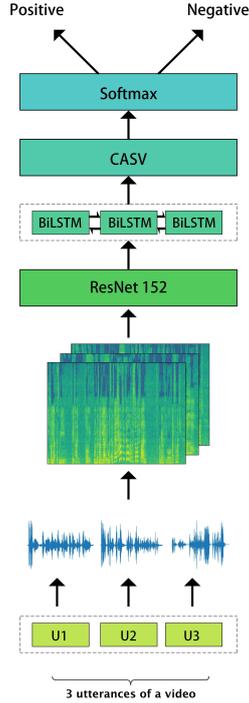} 
\caption{Overview of Our ResNet152 CNN Model} 
\label{Fig.main3} 
\end{figure}

Similar to the UB-BiLSTM model proposed above, we extracted the spectrum graph of each utterance through the Librosa toolkit and use it as the input of our CNN branch. After a lot of experiments, we found that the audio feature vector learned by the ResNet152 network structure has the best effect on the final sentiment classification, so we choose the ResNet model in this branch. The convolutional layer performs 2-dimensional convolution between the spectrum graph and the predefined linear filters. To enable the network to extract complementary features and learn the characteristics of input spectrum graph, a number of filters with different functions are used. A more refined audio feature vector is obtained through deep convolutional neural network, and then put into the BiLSTM layer to learn related sentiment information between adjacent utterances. Finally, before the softmax layer, we get another effective vector, named as CASV, extracted by our CNN framework, as shown in Figure 3. The process is described in CNN branch procedure in Algorithm 1.

\subsubsection{Fusion Layers}
Through the LSTM and CNN branches proposed above, we can extract two refined audio sentiment vectors, LASV and CASV for each utterance. We use these two kinds of vectors in parallel as the input of BiLSTM in AFF-ACRNN model. While effectively learning the relevant sentiment information of adjacent utterance, we extract the Audio Sentiment Vector (ASV) that has the greatest influence on the sentiment classification in the three utterances through the action of the attention mechanism. Finally, the final sentiment classification result is obtained by softmax layer. In \cite{donahue2015long}, Long-Term Recurrent Convolution Network (LRCN) model was proposed for visual recognition. LRCN is a consecutive structure of CNN and LSTM. LRCN processes the variable-length input with a CNN, whose outputs are fed into LSTM network, which finally predicts the class of the input. In \cite{sainath2015convolutional}, a cascade structure was used for voice search. Compared to the method mentioned above, the proposed network forms a parallel structure in which LSTM and CNN accept different inputs separately. Therefore, the Audio Sentiment Vector (ASV) can be extracted more comprehensively, and a relatively better classification result can be got.

\subsection{Feature Fusion base on Attention Mechanism} 
Inspired by human visual attention, the attention mechanism is proposed by \cite{bahdanau2014neural} in machine translation, which is introduced into the Encoder-Decoder framework to select the reference words in source language for words in target language. We use the attention mechanism to preserve the intermediate output of the input sequence by retaining the LSTM encoder, and then a model is trained to selectively learn these inputs and to correlate the output sequences with the model output. Specifically, when we fuse the features, each phoneme of the output sequence is associated with some specific frames in the input speech sequence, so that the feature representation that has the greatest influence on the final sentiment classification can be obtained, and finally obtain a fused Audio Feature Vector.  At the same time, attention mechanism behaves like a regulator since it can judge the importance of the contribution by adjacent relevant utterances for classifying the target utterance. Indeed, it is very hard to tell the sentiment of a single utterance if you do not concern its contextual information. However, you will also make a wrong estimation if contextual information is overly concerned. More specifically, in Figure 2, let $A_x$ be the $X_{th}$ attention network for utterance $U_x$, the corresponding attention weight vector is $\alpha_x$\, weighted hidden representation is $R_x$, we have:
\begin{equation}
{P_x} = tanh ({W_h}[x] \cdot H)
\end{equation}

\begin{equation}
{A_x} = softmax (w{[x]^T} \cdot {P_x})
\end{equation}

\begin{equation}
{R_x} = H \cdot {\alpha _x}^T
\end{equation}
Final representation for $x_th$ utterance is:
\begin{equation}
{h_x}^* = \tanh ({W_m}[x] \cdot {R_x} + {W_n}[x] \cdot {h_x})
\end{equation}
Where ${W_m}[x]$ and ${W_n}[x]$ are weights to be learned while training.

\section{Experiments}

In this section, we exhibit our experimental results and the analysis of our proposed model. More specifically, our model is trained and evaluated on utterance-level audio from CMU-MOSI dataset \cite{zadeh2016mosi} and being tested on MOUD \cite{perez2013utterance}. What's more, in order to verify that our proposed model has a strong generalization ability, we also carry out extensive expansion experiments.

\subsection{Experiment Setting}
\textbf{Evaluation Metrics} We evaluate our performance by weighted accuracy on both 2-class, 5-class and 7-class classification. 
\begin{equation}
weighted{\kern 1pt} {\kern 1pt} {\kern 1pt} {\kern 1pt} accuracy = \frac{{correct{\kern 1pt} {\kern 1pt} {\kern 1pt} {\kern 1pt} utterances}}{{utterances}}
\end{equation}
Additionally, F-Score is used to evaluate 2-class classification.

\begin{equation}
{{\rm{F}}_\beta }{\rm{ = }}(1 + {\beta ^2}) \cdot \frac{{precision \cdot recall}}{{({\beta ^2} \cdot precision) + recall}}
\end{equation}
where $\beta$ represents the weight between precision and recall. During our evaluation process, we set $\beta$ = 1 since we regard precision and recall has the same weight thus $F1$-score is adopted.

However, in 5-class and 7-class classification, we use Macro $F1$-Score to evaluate the result.
\begin{equation}
Macro{\kern 1pt} {\kern 1pt} {\kern 1pt} {\kern 1pt} {F_1}{\rm{ = }}\frac{{\sum\limits_1^n {{F_{1n}}} }}{n}
\end{equation}
where n represents the number of classification and $F_{1n}$ is the $F1$ score on $n{^{th}}$ category.

\subsubsection{Dataset details}
 CMU-MOSI dataset is rich in sentiment expressions, consisting 2199 opinionated utterances, 93 videos by 89 speakers. The videos address a large array of topics, such as movies, books, and products. Videos were crawled from YouTube and segmented into utterances where each utterance is annotated with scores between $-3$ (strongly negative) and +3 (strongly positive) by five annotators. We took the average of these five annotations as the sentiment polarity and considered three conditions where consists of two classes (positive and negative), five classes (strongly positive, positive, neutral, negative and strongly negative) and seven classes (strongly positive, positive, weakly positive, neutral, strongly negative,negative and weakly negative). Our train/test splits of the dataset are completely disjoint with respect to speakers. In order to better compare with the previous work, similar to \cite{poria2017context}, we divide the data set by 7:3 approximately, 1616 and 583 utterances are used for training and testing respectively. Furthermore, in order to verify that our model will not be heavily dependent on the language category, we tested it with the Spanish dataset MOUD. MOUD contains product review videos provided by 55 persons. The reviews are in Spanish.  The detailed datasets setup is depicted at Table 1.

\begin{table}[tp]
 
  \centering
  \fontsize{9}{12}\selectfont
  \begin{threeparttable}
  
  \label{tab:performance_comparison}
    \begin{tabular}{p{2cm} p{1cm}<{\centering} p{1cm}<{\centering} p{1cm}<{\centering} p{1cm}<{\centering}}
    \toprule
    \multirow{2}{*}{Datasets}&
    \multicolumn{2}{c}{ Train}&\multicolumn{2}{c}{ Test}\cr
    \cline{2-5}
    &utterance&video&utterance&video\cr
    \hline
    MOSI&1616&74&583&19\cr
    MOSI$\rightarrow$MOUD &2199&93&437&79\cr
    \bottomrule
    \end{tabular}
    \end{threeparttable}
    \caption{Datasets Setting. The right arrow means the model is trained and validated on the MOSI and tested on the MOUD}
\end{table}

\subsubsection{Network structure parameter}

Our proposed architecture is implemented based on the open-source deep learning framework Keras. More specifically, for proposed UB-BiLSTM framework, after a lot of experiments, we extracted the most four representative audio features of each utterance in a video through Librosa toolkit, which are MFCC, spectral\_centroid, chroma\_stft and spectral\_contrast respectively. In data processing, we make each utterance one-to-one correspondence with the label and rename the utterance. Accordingly, we extend each utterance to a feature matrix of $256*33$ dimensions. The output dimension of the first layer of BiLSTM is 128, and the second layer is 32. The output dimension of the first layer of Dense is 200, and the second is 2.

For the proposed CNN framework, the input images are warped into a fixed size of $512 * 512$. If the bounding box of the training samples provided, we firstly crop the images and then warp them to the fixed size. To train the feature encoder, we follow the fine-tuning training strategy. 

In all experiments, our networks are trained by Adam or SGD optimizer.Model is implemented by keras and GPU version is GeForce GTX TITAN X. In the LSTM branch, we initiate the learning rate to be 0.0001, and there are 200 epochs in the training part with batch size equals to 30 in each epoch. In the CNN branch, we initiate the learning rate to be 0.001, and there are 200 epochs in training Resnet-152 with batch size equals to 20 in each epoch.

Learning rate is short for lr and adam, sgd are the optimizers in keras.
Some major final settings for binary classifications on MOSI are: 

Lenet, batchsize =30, lr=0.01, epoch =120, adam

AlexNet, batchsize = 30, lr=0.001, epoch=100, sgd

VGG16, batchsize=30, lr=0.01, epoch=120, sgd

ZFNet batchsize=25, lr=0.001, epoch=100, sgd

Resnet18/50/152 batchsize = 20, lr=0.001, epoch =150/180/200, sgd

\begin{table}[ht]
\setlength{\belowcaptionskip}{10pt}
\centering
\begin{tabular}{llccc}
\hline
\multirow{2}{*}{\begin{tabular}[c]{@{}l@{}}Best Feature\\ Combination\end{tabular}} & \multirow{2}{*}{Model} & \multicolumn{3}{c}{Accuracy(\%)}                                                                        \\ \cline{3-5} 
                                                                                    &                        & \multicolumn{1}{l}{2-class}        & \multicolumn{1}{l}{5-class}        & \multicolumn{1}{l}{7-class} \\ \hline
\multirow{2}{*}{Single Type}                                                        & LSTM                   & \multicolumn{1}{c}{55.12}          & \multicolumn{1}{c}{23.64}          & 16.99                       \\
                                                                                    & BiLSTM                 & \multicolumn{1}{c}{55.98}          & \multicolumn{1}{c}{23.75}          & 17.24                       \\ 
\multirow{2}{*}{Two Types}                                                          & LSTM                   & \multicolumn{1}{c}{62.26}          & \multicolumn{1}{c}{28.23}          & 21.54                       \\
                                                                                    & BiLSTM                 & \multicolumn{1}{c}{63.76}          & \multicolumn{1}{c}{29.77}          & 22.92                       \\ 
\multirow{2}{*}{Thress Types}                                                       & LSTM                   & \multicolumn{1}{c}{66.36}          & \multicolumn{1}{c}{32.98}          & 24.66                       \\
                                                                                    & BiLSTM                 & \multicolumn{1}{c}{67.02}          & \multicolumn{1}{c}{33.75}          & 25.80                       \\ 
\multirow{2}{*}{\textbf{Four Types}}                                                & \textbf{LSTM}          & \multicolumn{1}{c}{\textbf{68.23}} & \multicolumn{1}{c}{\textbf{33.15}} & \textbf{26.27}              \\
                                                                                    & \textbf{BiLSTM}        & \multicolumn{1}{c}{\textbf{68.72}} & \multicolumn{1}{c}{\textbf{34.27}} & \textbf{26.82}              \\ 
\multirow{2}{*}{Five Types}                                                         & LSTM                   & \multicolumn{1}{c}{67.86}          & \multicolumn{1}{c}{31.29}          & 25.79                       \\
                                                                                    & BiLSTM                 & \multicolumn{1}{c}{67.97}          & \multicolumn{1}{c}{32.66}          & 26.01                       \\ 
\multirow{2}{*}{Six Types}                                                          & LSTM                   & \multicolumn{1}{c}{67.88}          & \multicolumn{1}{c}{32.23}          & 26.07                       \\
                                                                                    & BiLSTM                 & \multicolumn{1}{c}{68.61}          & \multicolumn{1}{c}{33.97}          & 26.78                       \\ 
\multirow{2}{*}{Seven Types}                                                        & LSTM                   & \multicolumn{1}{c}{68.01}          & \multicolumn{1}{c}{33.06}          & 25.99                       \\
                                                                                    & BiLSTM                 & \multicolumn{1}{c}{68.67}          & \multicolumn{1}{c}{34.18}          & 26.12                       \\ \hline
\end{tabular}
\caption{Comparison of different feature combinations}
\end{table}

\subsection{Performance Comparison}

\begin{table*}[ht]
\setlength{\belowcaptionskip}{10pt}

\centering
\begin{tabular}{cllcccccc}
\hline
\multicolumn{3}{c}{\multirow{2}{*}{Methods}} & \multicolumn{2}{c}{2-class} & \multicolumn{2}{c}{5-class} & \multicolumn{2}{c}{7-class}             \\ \cline{4-9} 
\multicolumn{3}{c}{}                         & Acc(\%)        & F1          & Acc(\%)      & Macro F1      & \multicolumn{1}{c}{Acc(\%)} & Macro F1 \\ \hline
\multicolumn{3}{c}{LeNet}                    & 56.75          & 55.62       & 23.67        & 21.87         & \multicolumn{1}{c}{15.63}   & 15.12    \\
\multicolumn{3}{c}{AlexNet}                  & 58.71          & 57.88       & 26.43        & 23.19         & \multicolumn{1}{c}{19.21}   & 18.79    \\
\multicolumn{3}{c}{VGG16}                    & 57.88          & 55.97       & 27.37        & 25.78         & \multicolumn{1}{c}{17.34}   & 16.25    \\
\multicolumn{3}{c}{ZFNet}                    & 55.37          & 53.12       & 21.90        & 21.38         & \multicolumn{1}{c}{12.82}   & 11.80    \\
\multicolumn{3}{c}{ResNet18}                 & 58.94          & 56.79       & 25.26        & 24.63         & \multicolumn{1}{c}{18.35}   & 17.89    \\
\multicolumn{3}{c}{ResNet50}                 & 62.52          & 61.21       & 28.13        & 27.04         & \multicolumn{1}{c}{20.21}   & 20.01    \\
\multicolumn{3}{c}{\textbf{ResNet152}}                & \textbf{65.42}          & \textbf{64.86}       & \textbf{28.78}       & \textbf{28.08}         & \multicolumn{1}{c}{\textbf{21.56}}   & \textbf{20.57}   \\ \hline
\end{tabular}
\caption{Comparison of SBCNN with different structure}
\end{table*}

\begin{table*}[ht]
\setlength{\belowcaptionskip}{10pt}

\centering
\begin{tabular}{cllcccccc}
\hline
\multicolumn{3}{c}{\multirow{2}{*}{Methods}} & \multicolumn{2}{c}{2-class}    & \multicolumn{2}{c}{5-class}    & \multicolumn{2}{c}{7-class}                          \\ \cline{4-9} 
\multicolumn{3}{c}{}                         & Acc(\%)        & F1             & Acc(\%)        & Macro F1       & \multicolumn{1}{c}{Acc(\%)}        & Macro F1       \\ \hline
\multicolumn{3}{c}{UB-Res18}            & 67.19          & 66.37          & 33.83          & 31.97          & \multicolumn{1}{c}{26.78}          & 25.83          \\
\multicolumn{3}{c}{UB-Res50}            & 67.83          & 66.69          & 34.21          & 33.78          & \multicolumn{1}{c}{27.75}          & 26.41          \\
\multicolumn{3}{c}{UB-Res152}           & 68.64          & 67.94          & 35.87          & 34.11          & \multicolumn{1}{c}{28.15}          & 27.03          \\
\multicolumn{3}{l}{UBBi-Res18}          & 68.26          & 66.25          & 35.43          & 33.52          & \multicolumn{1}{c}{27.63}          & 26.09          \\
\multicolumn{3}{l}{UBBi-Res50}          & 69.18          & 68.22          & 36.93          & 34.67          & \multicolumn{1}{c}{28.11}          & 27.54          \\
\multicolumn{3}{l}{\textbf{UBBi-Res152}}         & \textbf{69.64} & \textbf{68.51} & \textbf{37.71} & \textbf{35.12} & \multicolumn{1}{c}{\textbf{29.26}} & \textbf{28.45} \\ \hline
\end{tabular}
\caption{Comparison of different combinations between SBCNN and UB-BiLSTM}
\end{table*}

\subsubsection{Comparison of different feature combinations.}
Firstly, we have considered seven types of acoustic features that can best represent an audio,which mainly includes MFCC, root-mean-square energy, spectral and tonal features. A lot of experiments have been done in order to get the best feature combinations with different models on three types of classification. In the binary classification, for example, in order to find which are the best four features among the seven, we have carried out $C_7^4$ sets of experiments and only the best result of any 4-combination is recorded in bold in Table 2. The complete experiment result can be found at Appendix Part at the end of the paper.

What's more, we have also compared the different performance of our LASV extracted from LSTM-based and BiLSTM-based fusion model. As the Table 2 shows, the performance of LASV that extracted from BiLSTM-based model behaves better, since the acoustic information behind may also have impact on the acoustic information previous. It can be seen that the best number of feature combination is four and those four features are MFCC, spectral\_centroid, spectral\_contrast and chroma\_stft. That means the other three features, which are root-mean-square energy, spectral\_contrast and tonal centroid may introduce some noises or misleading in our sentiment analysis since all seven types of features do not have the best result. 

\subsubsection{Comparison of several renowned CNN-based model.}
We have compared our CASV performance extracted from the spectral map with several genres of popular models of CNN and its variants: LeNet \cite{lecun1998gradient}, AlexNet \cite{krizhevsky2012imagenet}, VGG16 \cite{simonyan2014very}, ZFNet \cite{zeiler2014visualizing}, ResNet \cite{he2016deep}. The results are listed in Table 3. As the neural network goes deeper, more representative features can be got from the spectrum graph and that is why ResNet152 has the best performance. It is benefited from the residual unit which will guarantee the network will not degrade when the network goes deeper. 

\subsubsection{Comparison of different combinations between SBCNN and UB-Bilstm}
At last, we have performed fusion experiments between several best SBCNN and UB-BiLSTM and UB-LSTM. More accurately, we choose the three best SBCNN, which are ReSNet18, ResNet50 and ResNet152 to combine with the two kinds of utterance dependent LSTM. The best combination is UB-BiLSTM with Res152. The final result is shown in Table 4.

\subsubsection{Comparison with traditional method.}
Apart from training deep neural network, a bunch of traditional binary classifiers has been used for sentiment analysis. In order to demonstrate the effectiveness of our model, we firstly compare our model with those traditional methods.

\cite{maghilnan2017sentiment} introduced a text-based SVM and Naive Bayes model for binary sentiment classification, thus we test their model on MOUD, rather than MOSI, to make comparison with our model because MOUD has only two sentiment level and each utterance has text record in the dataset.

\cite{bakir2018institutional} In this paper, except for SVM, the feature vectors like Mel Frequency Discrete Wavelet Coefficients (MFDWC), MFCC and Linear Predictive Cepstral Coefficients (LPCC) extracted from original record signal are trained with classification algorithm such as Dynamic Time Warping (DTW), Hidden Markov Model (HMM) and Gauss Mixture Model (GMM).

As shown in Table 5, we use weighted accuracy (ACC) and F1-Score to evaluate our results. Especially, for the ACC on MOUD, our proposed model outperforms the best model, SVM classifier, by 11.51\%.

\begin{table}[ht]
\centering
\setlength{\belowcaptionskip}{10pt}

\begin{tabular}{ccc}
\hline
\multirow{2}{*}{Model} & \multicolumn{2}{c}{MOUD}       \\ \cline{2-3} 
                       & ACC(\%)        & F1             \\ \hline
SVM                    & 57.23          & 54.83          \\
Naive Bayes            & 55.72          & 52.14          \\
GMM                    & 54.66          & 52.89          \\
HMM                    & 56.63          & 55.84          \\
DTW                    & 53.92          & 53.06          \\
\textbf{AFF-ACRNN}     & \textbf{68.74} & \textbf{66.37} \\ \hline
\end{tabular}
\caption{Comparison with traditional methods on MOUD}
\end{table}

\subsubsection{Comparison with the state-of-art.}
\cite{poria2017context} has introduced a LSTM-based model to utilize the contextual information extracted form each utterance in an video.
However, the input of the neural network model only has one type of feature, which is MFCC. This means all the utterance information is merely represented by one single feature. The acoustic information contained by the feature is somewhat duplicated and is bound to omit much sentiment information that might be hidden in many other useful features. What's worse, one type of feature means the input vector should be large enough to make sure that it carries enough information before it is fed into the neural network. This will undoubtedly increase the parameters to be trained in the network and meanwhile, it is time consuming and computation costly.

Our proposed model not only extracts the feature or sentiment vector from four types of traditional recognized acoustic features, have considered utterance dependency, but also extracts the feature from the spectrum graph, which may reveal some sentiment information that acoustic features cannot reflect. The final AFF-ACRNN consists of the  best combination of SBCNN and UB-BiLSTM and outperforms the state-of-the-art approach by 9.33\% in binary classification on MOSI dataset and by 12.54\% on MOUD. The results are shown in Table 6.

\begin{table}[ht]
\centering
\setlength{\belowcaptionskip}{10pt}

\begin{tabular}{ccc}
\hline
\multirow{2}{*}{Model} & \multicolumn{2}{c}{ACC(\%)}     \\ \cline{2-3} 
                       & MOSI   \quad     $\rightarrow$        & MOUD           \\ \hline
State-of-the-art       & \multicolumn{1}{l}{60.31}          & 47.20          \\
\textbf{AFF-ACRNN}     & \multicolumn{1}{l}{\textbf{69.64}} & \textbf{59.74} \\  \hline
\end{tabular}
\caption{Comparison with state-of-art result (Poria et al.2017) . The right arrow means the model is trained and validated on the MOSI and tested on the MOUD}
\end{table}

We have also run our model on one audio whose length is 10s for 1000 times and the average time to get the sentiment calssification result from input is only 655.94ms which thanks to our concentrated ASV extracted from AFF-ACRNN.

\subsection{Experiment Expansion}
Furthermore, considering the impact that different languages may have on the generalization capabilities of the proposed model, we experimented with the three largest language-related datasets in the world, which are Chinese, English, and Spanish. As mentioned above, CMU-MOSI is a English dataset, which is rigorously annotated with labels for subjectivity, sentiment intensity, per-frame and per-opinion annotated visual features, and per-milliseconds annotated audio features, but in this paper we only discuss annotated audio features. MOUD is a Spanish dataset, which collects a set of videos from the social media web site YouTube, using several keywords likely to lead to a product review or recommendation. It is worth mentioning that,  although Chinese is the most used language in the world, there is currently no public dataset available for experimentation. In this paper, our another contribution is to collate three different Chinese datasets. As is known to us, Chinese is wide-ranging and profound. Therefore, we compiled the three most representative Chinese datasets, which are Mandarin, Cantonese and Sichuan respectively. The detailed datasets setup is depicted at Table 7.

\begin{table}[tp]
 
  \centering
  \fontsize{9}{12}\selectfont
  \begin{threeparttable}
  
  \label{tab:performance_comparison}
    \begin{tabular}{p{2cm} p{1cm}<{\centering} p{1cm}<{\centering} p{1cm}<{\centering} p{1cm}<{\centering}}
    \toprule
    \multirow{2}{*}{Datasets}&
    \multicolumn{2}{c}{ Train}&\multicolumn{2}{c}{ Test}\cr
    \cline{2-5}
    &utterance&video&utterance&video\cr
    \hline
    Sichuan&1734&11&434&3\cr
    Cantonese&4123&36&1031&9\cr
    Mandarin&19764&75&4941&19\cr

    \bottomrule
    \end{tabular}
    \end{threeparttable}
    \caption{Expansion Experiment Datasets Setting}
\end{table}

Our Chinese datasets come from major online social media platforms or live video sites, including \href{https://weibo.com/}{Weibo}, \href{https://www.bilibili.com/}{bilibili}, \href{https://www.tiktok.com/}{Tik tok}  and so on. The content covers product reviews, movie reviews, shopping feedback and many other aspects of daily life. At the same time, taking into account the complexity of human emotions in reality and the differences in individual emotions, we draw on the annotation method of the CMU-MOSI dataset to find five people with psychology-related professional backgrounds to independently mark the scores, and finally 5 annotations. The scores are averaged to give the final emotional score. Finally, we divide the Chinese dataset into three categories, which are positive, neutral, and negative respectively. The detailed experiment results are shown in Table 8 and accuracy and loss of Chinese Mandarin training dataset and validation dataset is shown in Figure 4.

\begin{table}[ht]
\setlength{\belowcaptionskip}{10pt}
\centering
\begin{tabular}{llccc}
\hline
\multirow{2}{*}{\begin{tabular}[c]{@{}l@{}}Datasets\end{tabular}} & \multirow{2}{*}{Model} & \multicolumn{3}{c}{Accuracy(\%)}                                                                        \\ \cline{3-5} 
                                                                                    &                        & \multicolumn{1}{l}{Train}        & \multicolumn{1}{l}{Dev}        & \multicolumn{1}{l}{Test} \\ \hline
\multirow{1}{*}{Sichuan}                                                        
                                                                                    & AFF-ACRNN                 & \multicolumn{1}{c}{60.41}          & \multicolumn{1}{c}{57.11}          & 56.25                       \\ 
\multirow{1}{*}{Cantonese}                                                          
                                                                                    & AFF-ACRNN                 & \multicolumn{1}{c}{66.29}          & \multicolumn{1}{c}{61.72}          & 60.31                       \\ 
\multirow{1}{*}{Mandarin}                                                       
                                                                                    & AFF-ACRNN                 & \multicolumn{1}{c}{68.84}          & \multicolumn{1}{c}{64.10}          & 64.08                      \\ \hline

\end{tabular}
\caption{Expansion Experiment on Sichuan, Cantonese and Mandarin Datasets}
\end{table}

\begin{figure}[htbp] 
\centering
\includegraphics[width=0.6\textwidth]{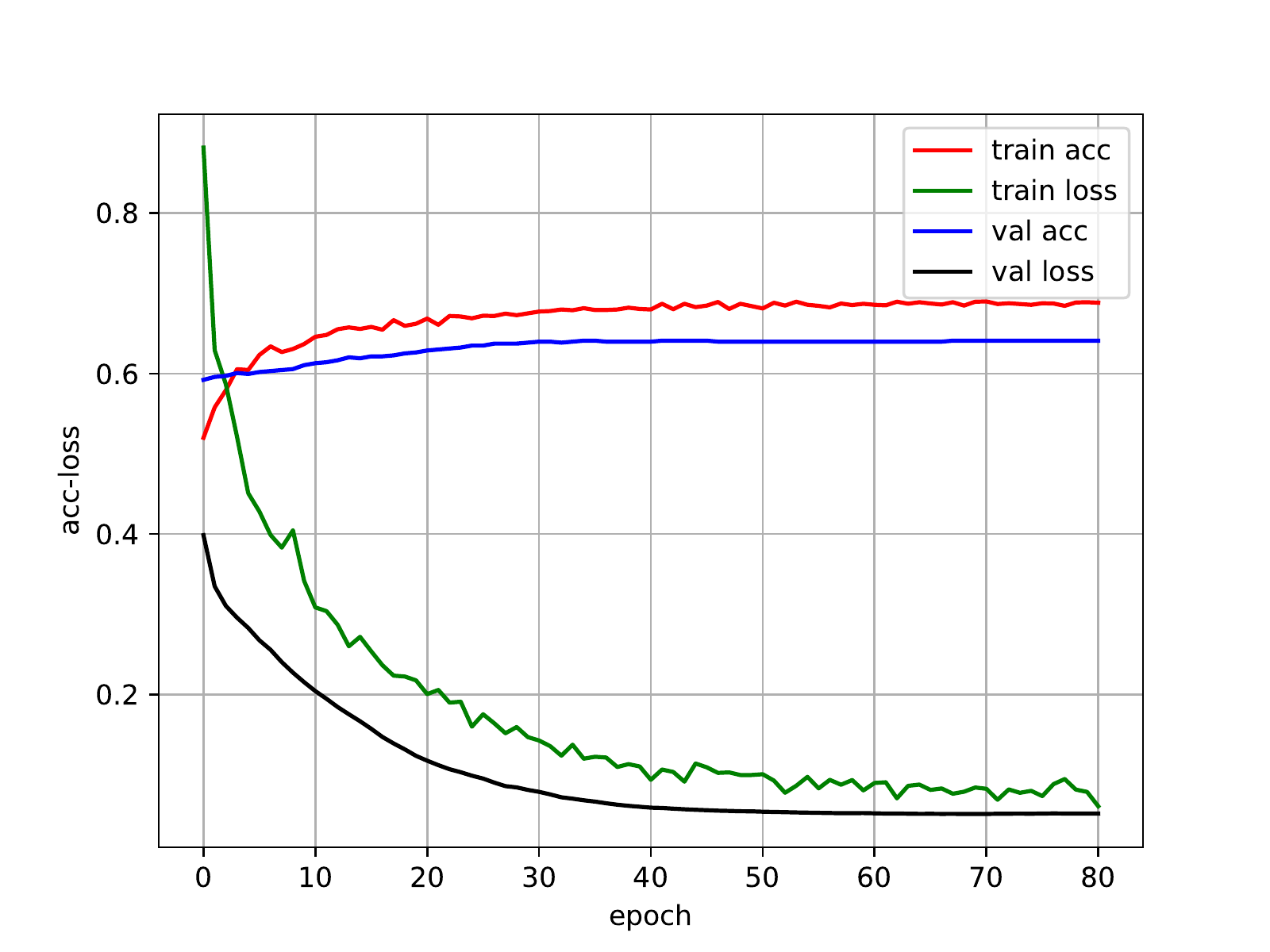} 
\caption{Experiment Result on Chinese Mandarin Dataset} 
\label{Fig.main3} 
\end{figure}

\subsection{Discussion}
The above experimental results have already shown us that the proposed method has a great improvement in the performance of audio sentiment analysis. In order to get the best structure of our AFF-ACRNN model, we have tested two separate branches respectively, and compare the final AFF-ACRNN with traditional or state-of-art method. Weighted accuracy and F1-Score, Macro F1-Score are used as metrics to evaluate the model's performance. In the UB-Bilstm branch, a lot of experiments have shown that four types of heterogeneous traditional features trained by BiLSTM will have the best result, whose weighted accuracy is 68.72\% on MOSI.  In the SBCNN branch, we have carried out seven experiments to prove the ResNet152 used in SBCNN will have the best result, for instance, with the weighted accuracy of 65.42\% on MOSI, due to its extreme depth and the helpful residual units used to prevent degradation. We selected six best combinations of SBCNN and UB-BiLSTM and find that the best is ResNet152 used in SBCNN with UB-Bilstm, whose weighted accuracy is 69.42\% on MOSI and outperforms not only the traditional classifier like SVM, but also the state-of-the-art approach by 9.33\% on MOSI dataset. Attention mechanism is used in both branch to subtly combine the heterogeneous acoustic features and choose the feature vectors that have the greatest impact on the sentiment classification. Furthermore, in the experiment of using MOSI as training set and verification set and MOUD as test set, it also shows that our proposed model has strong generalization ability.

\section{Conclusion and Future Work}

In this paper, we propose a novel utterance-based deep neural network model termed AFF-ACRNN, which has a parallel combination of CNN and LSTM based network, to obtain representative features termed ASV, that can maximally reflect sentiment information in an utterance from an audio. We extract several traditional heterogeneous acoustic features by Librosa toolkit and choose the four most representative features through a large number of experiments, and regard them as the input of the neural network. We can get CASV and LASV from the CNN branch and the LSTM branch respectively, and finally merge the two branches to obtain the final ASV for sentiment classification of each utterance. Besides, BiLSTM with attention mechanism is used for feature fusion. The experiment results show our model can recognize audio sentiment precisely and quickly, and demonstrate our heterogeneous ASV are better than traditional acoustic features or vectors extracted from other deep learning models. Furthermore, experiment results indicate that the proposed model outperforms the state-of-the-art approach by 9.33\% on MOSI dataset. We have also tested our model on MOUD to prove the model won't heavily depend on language types. In the future, we will combine the feature engineering technologies to further discuss the fusion dimension of audio features and consider the fusion of different dimensions of different categories of features, and even apply them to multimodal sentiment analysis.


\section*{References}

\bibliography{mybibfile}

\newpage
\section*{Appendix}
\vspace{2cm}

\begin{table}[ht]
\centering
\begin{tabular}{lllll}
\hline
\multirow{2}{*}{Combinations of One Type of Feature}                   & \multirow{2}{*}{Model} & \multicolumn{3}{c}{Accuracy(\%)}                 \\ \cline{3-5} 
                                                                       &                        & 2-class        & 5-class        & 7-class        \\ \hline
\multirow{2}{*}{1 Chromagram from spectrogram (chroma\_stft)}          & LSTM                   & 43.24          & 20.23          & 13.96          \\
                                                                       & BiLSTM                 & 45.37          & 2.29           & 12.39          \\
\multirow{2}{*}{2 Chroma Energy Normalized (chroma\_cens)}             & LSTM                   & 42.98          & 20.87          & 13.31          \\
                                                                       & BiLSTM                 & 45.85          & 20.53          & 13.76          \\
\multirow{2}{*}{\textbf{3 Mel-frequency cepstral coefficients (MFCC)}} & \textbf{LSTM}          & \textbf{55.12} & \textbf{23.64} & \textbf{16.99} \\
                                                                       & \textbf{BiLSTM}        & \textbf{55.98} & \textbf{23.75} & \textbf{17.24} \\
\multirow{2}{*}{4 Root-Mean-Square Energy (RMSE)}                      & LSTM                   & 52.30          & 21.14          & 15.33          \\
                                                                       & BiLSTM                 & 52.76          & 22.35          & 15.87          \\
\multirow{2}{*}{5 Spectral\_Centroid}                                  & LSTM                   & 48.39          & 22.25          & 14.97          \\
                                                                       & BiLSTM                 & 48.84          & 22.36          & 15.79          \\
\multirow{2}{*}{6 Spectral\_Contrast}                                  & LSTM                   & 48.34          & 22.50          & 15.02          \\
                                                                       & BiLSTM                 & 48.97          & 22.28          & 15.98          \\
\multirow{2}{*}{7 Tonal Centroid Features (tonnetz)}                   & LSTM                   & 53.78          & 22.67          & 15.83          \\
                                                                       & BiLSTM                 & 54.24          & 21.87          & 16.01          \\ \hline
\end{tabular}
\caption{Comparison of One Type of Feature}
\end{table}

\vspace{1cm}
Numbers in table 10-15 have the following correspondence :  \\[0.5cm]
1 : Chromagram from spectrogram (chroma\_stft) \\[0.5cm]
2 : Chroma Energy Normalized (chroma\_cens) \\[0.5cm]
3 : Mel-frequency cepstral coefficients (MFCC) \\[0.5cm]
4 : Root-Mean-Square Energy (RMSE) \\[0.5cm]
5 : Spectral\_Centroid \\[0.5cm]
6 : Spectral\_Contrast \\[0.5cm]
7 : Tonal Centroid Features (tonnetz)

\begin{table}[ht]
\centering
\begin{tabular}{lllll}
\hline
\multirow{2}{*}{Combinations of Two Types of Features} & \multirow{2}{*}{Model} & \multicolumn{3}{c}{Accuracy(\%)}                 \\ \cline{3-5} 
                                                       &                        & 2-class        & 5-class        & 7-class        \\ \hline
\multirow{2}{*}{1\&2}                                  & LSTM                   & 59.85          & 24.82          & 18.62          \\
                                                       & BiLSTM                 & 60.42          & 25.73          & 19.42          \\
\multirow{2}{*}{1\&3}                                  & LSTM                   & 61.66          & 27.73          & 20.55          \\
                                                       & BiLSTM                 & 62.72          & 28.81          & 21.31          \\
\multirow{2}{*}{1\&4}                                  & LSTM                   & 59.42          & 24.13          & 18.09          \\
                                                       & BiLSTM                 & 60.23          & 25.29          & 19.17          \\
\multirow{2}{*}{1\&5}                                  & LSTM                   & 60.03          & 25.02          & 19.03          \\
                                                       & BiLSTM                 & 61.21          & 25.93          & 20.14          \\
\multirow{2}{*}{1\&6}                                  & LSTM                   & 60.18          & 25.17          & 19.23          \\
                                                       & BiLSTM                 & 61.42          & 26.05          & 20.38          \\
\multirow{2}{*}{1\&7}                                  & LSTM                   & 59.35          & 24.78          & 18.53          \\
                                                       & BiLSTM                 & 60.13          & 25.35          & 19.28          \\
\multirow{2}{*}{2\&3}                                  & LSTM                   & 61.24          & 27.32          & 20.64          \\
                                                       & BiLSTM                 & 62.38          & 28.63          & 21.86          \\
\multirow{2}{*}{2\&4}                                  & LSTM                   & 58.82          & 24.63          & 18.12          \\
                                                       & BiLSTM                 & 60.25          & 25.21          & 18.79          \\
\multirow{2}{*}{2\&5}                                  & LSTM                   & 59.23          & 24.05          & 18.15          \\
                                                       & BiLSTM                 & 60.53          & 25.32          & 19.27          \\
\multirow{2}{*}{2\&6}                                  & LSTM                   & 59.32          & 24.48          & 18.26          \\
                                                       & BiLSTM                 & 60.83          & 25.58          & 19.46          \\
\multirow{2}{*}{2\&7}                                  & LSTM                   & 58.82          & 23.62          & 17.93          \\
                                                       & BiLSTM                 & 58.95          & 24.84          & 18.35          \\
\multirow{2}{*}{3\&4}                                  & LSTM                   & 62.03          & 27.86          & 21.25          \\
                                                       & BiLSTM                 & 62.87          & 29.05          & 22.36          \\
\multirow{2}{*}{\textbf{3\&5}}                         & \textbf{LSTM}          & \textbf{62.26} & \textbf{28.23} & \textbf{21.54} \\
                                                       & \textbf{BiLSTM}        & \textbf{63.76} & \textbf{29.77} & \textbf{22.92} \\
\multirow{2}{*}{3\&6}                                  & LSTM                   & 62.19          & 28.08          & 21.36          \\
                                                       & BiLSTM                 & 63.66          & 29.72          & 22.83          \\
\multirow{2}{*}{3\&7}                                  & LSTM                   & 61.92          & 27.89          & 20.96          \\
                                                       & BiLSTM                 & 62.38          & 28.83          & 22.62          \\
\multirow{2}{*}{4\&5}                                  & LSTM                   & 59.04          & 24.45          & 18.42          \\
                                                       & BiLSTM                 & 60.64          & 25.55          & 19.37          \\
\multirow{2}{*}{4\&6}                                  & LSTM                   & 59.02          & 24.25          & 18.37          \\
                                                       & BiLSTM                 & 60.48          & 25.27          & 19.28          \\
\multirow{2}{*}{4\&7}                                  & LSTM                   & 58.23          & 23.35          & 17.35          \\
                                                       & BiLSTM                 & 59.92          & 22.75          & 18.46          \\
\multirow{2}{*}{5\&6}                                  & LSTM                   & 61.31          & 27.29          & 20.84          \\
                                                       & BiLSTM                 & 62.46          & 28.47          & 21.52          \\
\multirow{2}{*}{5\&7}                                  & LSTM                   & 59.37          & 25.02          & 19.06          \\
                                                       & BiLSTM                 & 60.84          & 25.83          & 19.9           \\
\multirow{2}{*}{6\&7}                                  & LSTM                   & 59.21          & 24.79          & 18.83          \\
                                                       & BiLSTM                 & 60.71          & 25.52          & 19.72          \\ \hline
\end{tabular}
\caption{Comparison of Two Types of Feature}
\end{table}

\begin{table}[ht]
\centering
\scriptsize
\begin{tabular}{lllll}
\hline
\multirow{2}{*}{Combinations of Three Types of Features} & \multirow{2}{*}{Model} & \multicolumn{3}{c}{Accuracy(\%)}                 \\ \cline{3-5} 
                                                         &                        & 2-class        & 5-class        & 7-class        \\ \hline
\multirow{2}{*}{1\&2\&3}                                 & LSTM                   & 65.82          & 32.51          & 24.63          \\
                                                         & BiLSTM                 & 66.73          & 33.84          & 25.75          \\
\multirow{2}{*}{1\&2\&4}                                 & LSTM                   & 62.16          & 31.38          & 22.39          \\
                                                         & BiLSTM                 & 63.82          & 32.39          & 23.46          \\
\multirow{2}{*}{1\&2\&5}                                 & LSTM                   & 62.39          & 31.82          & 22.81          \\
                                                         & BiLSTM                 & 63.97          & 32.79          & 23.72          \\
\multirow{2}{*}{1\&2\&6}                                 & LSTM                   & 62.24          & 31.41          & 22.39          \\
                                                         & BiLSTM                 & 63.93          & 32.37          & 23.37          \\
\multirow{2}{*}{1\&2\&7}                                 & LSTM                   & 62.04          & 31.34          & 22.35          \\
                                                         & BiLSTM                 & 63.80          & 32.29          & 23.30          \\
\multirow{2}{*}{1\&3\&4}                                 & LSTM                   & 63.72          & 32.31          & 23.74          \\
                                                         & BiLSTM                 & 64.75          & 33.64          & 25.02          \\
\multirow{2}{*}{1\&3\&5}                                 & LSTM                   & 65.83          & 32.42          & 24.33          \\
                                                         & BiLSTM                 & 66.92          & 33.73          & 25.82          \\
\multirow{2}{*}{1\&3\&6}                                 & LSTM                   & 65.87          & 32.53          & 24.77          \\
                                                         & BiLSTM                 & 66.93          & 33.89          & 25.85          \\
\multirow{2}{*}{1\&3\&7}                                 & LSTM                   & 65.76          & 32.34          & 24.04          \\
                                                         & BiLSTM                 & 66.38          & 33.24          & 25.23          \\
\multirow{2}{*}{1\&4\&5}                                 & LSTM                   & 64.23          & 31.25          & 23.77          \\
                                                         & BiLSTM                 & 65.34          & 32.18          & 24.63          \\
\multirow{2}{*}{1\&4\&6}                                 & LSTM                   & 63.98          & 30.87          & 23.69          \\
                                                         & BiLSTM                 & 64.91          & 31.96          & 24.52          \\
\multirow{2}{*}{1\&4\&7}                                 & LSTM                   & 63.81          & 30.25          & 23.61          \\
                                                         & BiLSTM                 & 64.72          & 31.45          & 24.05          \\
\multirow{2}{*}{1\&5\&6}                                 & LSTM                   & 65.27          & 30.92          & 23.93          \\
                                                         & BiLSTM                 & 66.01          & 31.99          & 24.22          \\
\multirow{2}{*}{1\&5\&7}                                 & LSTM                   & 64.16          & 30.28          & 23.12          \\
                                                         & BiLSTM                 & 64.99          & 30.77          & 23.88          \\
\multirow{2}{*}{1\&6\&7}                                 & LSTM                   & 64.28          & 30.23          & 23.08          \\
                                                         & BiLSTM                 & 65.21          & 30.82          & 23.83          \\
\multirow{2}{*}{2\&3\&4}                                 & LSTM                   & 65.02          & 31.21          & 23.07          \\
                                                         & BiLSTM                 & 66.13          & 31.87          & 24.21          \\
\multirow{2}{*}{2\&3\&5}                                 & LSTM                   & 65.78          & 31.92          & 23.99          \\
                                                         & BiLSTM                 & 66.84          & 32.85          & 24.86          \\
\multirow{2}{*}{2\&3\&6}                                 & LSTM                   & 65.23          & 31.28          & 23.83          \\
                                                         & BiLSTM                 & 66.34          & 32.68          & 24.79          \\
\multirow{2}{*}{2\&3\&7}                                 & LSTM                   & 64.73          & 31.09          & 23.48          \\
                                                         & BiLSTM                 & 65.88          & 32.34          & 24.24          \\
\multirow{2}{*}{2\&4\&5}                                 & LSTM                   & 64.72          & 31.33          & 24.04          \\
                                                         & BiLSTM                 & 65.23          & 32.45          & 24.80          \\
\multirow{2}{*}{2\&4\&6}                                 & LSTM                   & 64.54          & 31.05          & 23.94          \\
                                                         & BiLSTM                 & 64.98          & 32.03          & 24.72          \\
\multirow{2}{*}{2\&4\&7}                                 & LSTM                   & 63.02          & 31.34          & 23.84          \\
                                                         & BiLSTM                 & 63.85          & 31.93          & 24.05          \\
\multirow{2}{*}{2\&5\&6}                                 & LSTM                   & 65.31          & 31.88          & 24.15          \\
                                                         & BiLSTM                 & 66.35          & 32.34          & 24.74          \\
\multirow{2}{*}{2\&5\&7}                                 & LSTM                   & 64.85          & 31.84          & 23.02          \\
                                                         & BiLSTM                 & 65.73          & 32.15          & 23.68          \\
\multirow{2}{*}{2\&6\&7}                                 & LSTM                   & 64.82          & 31.95          & 23.06          \\
                                                         & BiLSTM                 & 65.83          & 32.03          & 23.79          \\
\multirow{2}{*}{3\&4\&5}                                 & LSTM                   & 65.18          & 32.17          & 23.81          \\
                                                         & BiLSTM                 & 66.38          & 33.10          & 24.64          \\
\multirow{2}{*}{3\&4\&6}                                 & LSTM                   & 64.29          & 31.46          & 23.56          \\
                                                         & BiLSTM                 & 65.83          & 32.19          & 23.95          \\
\multirow{2}{*}{3\&4\&7}                                 & LSTM                   & 64.72          & 31.07          & 23.09          \\
                                                         & BiLSTM                 & 65.86          & 32.23          & 23.92          \\
\multirow{2}{*}{\textbf{3\&5\&6}}                        & \textbf{LSTM}          & \textbf{66.36} & \textbf{32.98} & \textbf{24.66} \\
                                                         & \textbf{BiLSTM}        & \textbf{67.02} & \textbf{33.75} & \textbf{25.80} \\
\multirow{2}{*}{3\&5\&7}                                 & LSTM                   & 65.83          & 32.76          & 24.24          \\
                                                         & BiLSTM                 & 66.74          & 33.56          & 25.62          \\
\multirow{2}{*}{3\&6\&7}                                 & LSTM                   & 65.78          & 32.62          & 24.03          \\
                                                         & BiLSTM                 & 66.61          & 33.36          & 25.15          \\
\multirow{2}{*}{4\&5\&6}                                 & LSTM                   & 65.31          & 31.64          & 24.28          \\
                                                         & BiLSTM                 & 66.29          & 32.19          & 24.89          \\
\multirow{2}{*}{4\&5\&7}                                 & LSTM                   & 64.58          & 31.36          & 24.02          \\
                                                         & BiLSTM                 & 64.92          & 32.07          & 24.85          \\
\multirow{2}{*}{4\&6\&7}                                 & LSTM                   & 64.55          & 31.29          & 23.99          \\
                                                         & BiLSTM                 & 64.87          & 32.08          & 24.76          \\
\multirow{2}{*}{5\&6\&7}                                 & LSTM                   & 65.24          & 31.58          & 24.08          \\
                                                         & BiLSTM                 & 66.15          & 32.17          & 24.71          \\ \hline
\end{tabular}
\caption{Comparison of Three Types of Feature}
\end{table}

\begin{table}[ht]
\centering
\scriptsize
\begin{tabular}{lllll}
\hline
\multirow{2}{*}{Combinations of Four Types of Features} & \multirow{2}{*}{Model} & \multicolumn{3}{c}{Accuracy(\%)}                 \\ \cline{3-5} 
                                                        &                        & 2-class        & 5-class        & 7-class        \\ \hline
\multirow{2}{*}{1\&2\&3\&4}                             & LSTM                   & 67.74          & 32.28          & 25.78          \\
                                                        & BiLSTM                 & 68.01          & 32.91          & 26.06          \\
\multirow{2}{*}{1\&2\&3\&5}                             & LSTM                   & 68.09          & 33.02          & 26.14          \\
                                                        & BiLSTM                 & 68.57          & 34.07          & 26.73          \\
\multirow{2}{*}{1\&2\&3\&6}                             & LSTM                   & 68.03          & 32.99          & 26.01          \\
                                                        & BiLSTM                 & 68.52          & 34.05          & 26.67          \\
\multirow{2}{*}{1\&2\&3\&7}                             & LSTM                   & 67.78          & 32.45          & 25.81          \\
                                                        & BiLSTM                 & 68.02          & 33.09          & 26.13          \\
\multirow{2}{*}{1\&2\&4\&5}                             & LSTM                   & 67.66          & 31.94          & 25.17          \\
                                                        & BiLSTM                 & 67.92          & 32.07          & 25.92          \\
\multirow{2}{*}{1\&2\&4\&6}                             & LSTM                   & 67.03          & 31.14          & 24.98          \\
                                                        & BiLSTM                 & 67.61          & 31.96          & 25.31          \\
\multirow{2}{*}{1\&2\&4\&7}                             & LSTM                   & 67.01          & 31.08          & 24.77          \\
                                                        & BiLSTM                 & 67.55          & 31.73          & 25.07          \\
\multirow{2}{*}{1\&2\&5\&6}                             & LSTM                   & 67.93          & 31.86          & 25.01          \\
                                                        & BiLSTM                 & 68.06          & 32.28          & 25.92          \\
\multirow{2}{*}{1\&2\&5\&7}                             & LSTM                   & 67.72          & 31.79          & 24.91          \\
                                                        & BiLSTM                 & 68.00          & 32.14          & 25.68          \\
\multirow{2}{*}{1\&2\&6\&7}                             & LSTM                   & 67.52          & 31.37          & 24.61          \\
                                                        & BiLSTM                 & 67.89          & 31.92          & 25.03          \\
\multirow{2}{*}{1\&3\&4\&5}                             & LSTM                   & 68.07          & 33.01          & 26.12          \\
                                                        & BiLSTM                 & 68.56          & 34.06          & 26.71          \\
\multirow{2}{*}{1\&3\&4\&6}                             & LSTM                   & 68.05          & 33.02          & 26.08          \\
                                                        & BiLSTM                 & 68.53          & 34.03          & 26.57          \\
\multirow{2}{*}{1\&3\&4\&7}                             & LSTM                   & 67.61          & 32.02          & 25.32          \\
                                                        & BiLSTM                 & 67.92          & 32.46          & 26.79          \\
\multirow{2}{*}{\textbf{1\&3\&5\&6}}                    & \textbf{LSTM}          & \textbf{68.23} & \textbf{33.15} & \textbf{26.27} \\
                                                        & \textbf{BiLSTM}        & \textbf{68.72} & \textbf{34.27} & \textbf{26.82} \\
\multirow{2}{*}{1\&3\&5\&7}                             & LSTM                   & 68.06          & 33.04          & 26.12          \\
                                                        & BiLSTM                 & 68.55          & 34.13          & 26.72          \\
\multirow{2}{*}{1\&3\&6\&7}                             & LSTM                   & 68.02          & 32.97          & 26.01          \\
                                                        & BiLSTM                 & 68.49          & 33.89          & 26.64          \\
\multirow{2}{*}{1\&4\&5\&6}                             & LSTM                   & 67.83          & 32.27          & 26.92          \\
                                                        & BiLSTM                 & 68.12          & 32.51          & 26.31          \\
\multirow{2}{*}{1\&4\&5\&7}                             & LSTM                   & 67.33          & 32.16          & 26.76          \\
                                                        & BiLSTM                 & 67.99          & 32.48          & 26.89          \\
\multirow{2}{*}{1\&4\&6\&7}                             & LSTM                   & 67.20          & 32.12          & 26.59          \\
                                                        & BiLSTM                 & 67.71          & 32.39          & 26.77          \\
\multirow{2}{*}{1\&5\&6\&7}                             & LSTM                   & 67.45          & 32.34          & 26.98          \\
                                                        & BiLSTM                 & 67.92          & 32.61          & 27.02          \\
\multirow{2}{*}{2\&3\&4\&5}                             & LSTM                   & 67.97          & 32.54          & 25.92          \\
                                                        & BiLSTM                 & 68.12          & 33.61          & 26.18          \\
\multirow{2}{*}{2\&3\&4\&6}                             & LSTM                   & 67.50          & 32.41          & 27.03          \\
                                                        & BiLSTM                 & 67.98          & 32.79          & 27.19          \\
\multirow{2}{*}{2\&3\&4\&7}                             & LSTM                   & 67.41          & 32.39          & 26.91          \\
                                                        & BiLSTM                 & 67.83          & 32.74          & 27.02          \\
\multirow{2}{*}{2\&3\&5\&6}                             & LSTM                   & 68.19          & 33.06          & 26.21          \\
                                                        & BiLSTM                 & 68.67          & 34.13          & 26.79          \\
\multirow{2}{*}{2\&3\&5\&7}                             & LSTM                   & 67.52          & 32.06          & 25.48          \\
                                                        & BiLSTM                 & 67.91          & 33.15          & 25.86          \\
\multirow{2}{*}{2\&3\&6\&7}                             & LSTM                   & 67.71          & 32.58          & 27.26          \\
                                                        & BiLSTM                 & 67.93          & 33.01          & 27.98          \\
\multirow{2}{*}{2\&4\&5\&6}                             & LSTM                   & 67.67          & 32.55          & 27.23          \\
                                                        & BiLSTM                 & 67.90          & 33.96          & 27.94          \\
\multirow{2}{*}{2\&4\&5\&7}                             & LSTM                   & 67.44          & 32.29          & 27.01          \\
                                                        & BiLSTM                 & 67.78          & 33.67          & 27.66          \\
\multirow{2}{*}{2\&4\&6\&7}                             & LSTM                   & 67.22          & 32.11          & 26.97          \\
                                                        & BiLSTM                 & 67.69          & 32.98          & 27.33          \\
\multirow{2}{*}{2\&5\&6\&7}                             & LSTM                   & 67.37          & 32.26          & 26.99          \\
                                                        & BiLSTM                 & 67.60          & 33.04          & 27.38          \\
\multirow{2}{*}{3\&4\&5\&6}                             & LSTM                   & 68.11          & 33.04          & 26.18          \\
                                                        & BiLSTM                 & 68.62          & 34.09          & 26.75          \\
\multirow{2}{*}{3\&4\&5\&7}                             & LSTM                   & 67.19          & 31.98          & 25.07          \\
                                                        & BiLSTM                 & 67.23          & 33.02          & 25.71          \\
\multirow{2}{*}{3\&4\&6\&7}                             & LSTM                   & 67.81          & 32.95          & 27.41          \\
                                                        & BiLSTM                 & 68.02          & 33.31          & 27.85          \\
\multirow{2}{*}{3\&5\&6\&7}                             & LSTM                   & 68.17          & 33.05          & 26.20          \\
                                                        & BiLSTM                 & 68.65          & 34.11          & 26.77          \\
\multirow{2}{*}{4\&5\&6\&7}                             & LSTM                   & 67.61          & 33.01          & 26.71          \\
                                                        & BiLSTM                 & 67.96          & 33.87          & 27.04          \\ \hline
\end{tabular}
\caption{Comparison of Four Types of Feature}
\end{table}

\begin{table}[ht]
\centering
\begin{tabular}{lllll}
\hline
\multirow{2}{*}{Combinations of Five Types of Features} & \multirow{2}{*}{Model} & \multicolumn{3}{c}{Accuracy(\%)}                 \\ \cline{3-5} 
                                                        &                        & 2-class        & 5-class        & 7-class        \\ \hline
\multirow{2}{*}{1\&2\&3\&4\&5}                          & LSTM                   & 67.58          & 30.62          & 25.01          \\
                                                        & BiLSTM                 & 67.69          & 31.19          & 25.63          \\
\multirow{2}{*}{1\&2\&3\&4\&6}                          & LSTM                   & 67.54          & 30.51          & 24.91          \\
                                                        & BiLSTM                 & 67.62          & 30.92          & 25.24          \\
\multirow{2}{*}{1\&2\&3\&4\&7}                          & LSTM                   & 67.51          & 30.33          & 24.82          \\
                                                        & BiLSTM                 & 67.60          & 30.79          & 25.19          \\
\multirow{2}{*}{1\&2\&3\&5\&6}                          & LSTM                   & 67.80          & 30.75          & 25.21          \\
                                                        & BiLSTM                 & 67.84          & 31.69          & 25.87          \\
\multirow{2}{*}{1\&2\&3\&5\&7}                          & LSTM                   & 67.55          & 30.59          & 24.92          \\
                                                        & BiLSTM                 & 67.63          & 30.98          & 25.39          \\
\multirow{2}{*}{1\&2\&3\&6\&7}                          & LSTM                   & 67.53          & 30.55          & 24.79          \\
                                                        & BiLSTM                 & 67.59          & 30.82          & 25.11          \\
\multirow{2}{*}{1\&2\&4\&5\&6}                          & LSTM                   & 67.43          & 30.38          & 24.51          \\
                                                        & BiLSTM                 & 67.56          & 30.78          & 24.98          \\
\multirow{2}{*}{1\&2\&4\&5\&7}                          & LSTM                   & 67.22          & 30.15          & 24.17          \\
                                                        & BiLSTM                 & 67.49          & 30.61          & 24.73          \\
\multirow{2}{*}{1\&2\&4\&6\&7}                          & LSTM                   & 66.92          & 29.87          & 23.91          \\
                                                        & BiLSTM                 & 67.05          & 30.24          & 24.11          \\
\multirow{2}{*}{1\&2\&5\&6\&7}                          & LSTM                   & 67.02          & 30.13          & 24.28          \\
                                                        & BiLSTM                 & 67.31          & 30.52          & 24.73          \\
\multirow{2}{*}{\textbf{1\&3\&4\&5\&6}}                 & \textbf{LSTM}          & \textbf{67.86} & \textbf{31.29} & \textbf{25.79} \\
                                                        & \textbf{BiLSTM}        & \textbf{67.97} & \textbf{32.66} & \textbf{26.01} \\
\multirow{2}{*}{1\&3\&4\&5\&7}                          & LSTM                   & 67.56          & 30.60          & 24.97          \\
                                                        & BiLSTM                 & 67.67          & 31.06          & 25.56          \\
\multirow{2}{*}{1\&3\&4\&6\&7}                          & LSTM                   & 67.39          & 30.43          & 24.93          \\
                                                        & BiLSTM                 & 67.52          & 30.99          & 25.37          \\
\multirow{2}{*}{1\&3\&5\&6\&7}                          & LSTM                   & 67.83          & 30.96          & 25.38          \\
                                                        & BiLSTM                 & 67.88          & 31.77          & 25.91          \\
\multirow{2}{*}{1\&4\&5\&6\&7}                          & LSTM                   & 66.87          & 29.73          & 23.77          \\
                                                        & BiLSTM                 & 67.01          & 30.02          & 24.05          \\
\multirow{2}{*}{2\&3\&4\&5\&6}                          & LSTM                   & 67.62          & 30.68          & 25.19          \\
                                                        & BiLSTM                 & 67.79          & 31.42          & 25.82          \\
\multirow{2}{*}{2\&3\&4\&5\&7}                          & LSTM                   & 67.53          & 30.42          & 24.89          \\
                                                        & BiLSTM                 & 67.61          & 30.87          & 25.17          \\
\multirow{2}{*}{2\&3\&4\&6\&7}                          & LSTM                   & 66.70          & 29.25          & 23.98          \\
                                                        & BiLSTM                 & 66.93          & 29.94          & 24.46          \\
\multirow{2}{*}{2\&3\&5\&6\&7}                          & LSTM                   & 67.60          & 30.65          & 25.07          \\
                                                        & BiLSTM                 & 67.73          & 31.23          & 25.66          \\
\multirow{2}{*}{2\&4\&5\&6\&7}                          & LSTM                   & 67.48          & 30.12          & 24.71          \\
                                                        & BiLSTM                 & 67.69          & 30.94          & 25.31          \\
\multirow{2}{*}{3\&4\&5\&6\&7}                          & LSTM                   & 67.64          & 30.71          & 25.27          \\
                                                        & BiLSTM                 & 67.81          & 31.58          & 25.94          \\ \hline
\end{tabular}
\caption{Comparison of Five Types of Feature}
\end{table}

\begin{table}[ht]
\centering
\begin{tabular}{lllll}
\hline
\multirow{2}{*}{Combinations of Six Types of Feature} & \multirow{2}{*}{Model} & \multicolumn{3}{c}{Accuracy(\%)}                 \\ \cline{3-5} 
                                                      &                        & 2-class        & 5-class        & 7-class        \\ \hline
\multirow{2}{*}{1\&2\&3\&4\&5\&6}                     & LSTM                   & 67.84          & 32.21          & 26.04          \\
                                                      & BiLSTM                 & 68.60          & 33.92          & 26.71          \\
\multirow{2}{*}{1\&2\&3\&4\&5\&7}                     & LSTM                   & 67.72          & 32.09          & 25.81          \\
                                                      & BiLSTM                 & 68.38          & 33.68          & 26.49          \\
\multirow{2}{*}{1\&2\&3\&4\&6\&7}                     & LSTM                   & 67.69          & 32.01          & 25.68          \\
                                                      & BiLSTM                 & 68.27          & 33.57          & 26.34          \\
\multirow{2}{*}{1\&2\&3\&5\&6\&7}                     & LSTM                   & 67.82          & 32.19          & 25.91          \\
                                                      & BiLSTM                 & 68.59          & 33.88          & 26.63          \\
\multirow{2}{*}{1\&2\&4\&5\&6\&7}                     & LSTM                   & 67.61          & 31.97          & 25.62          \\
                                                      & BiLSTM                 & 68.22          & 33.35          & 26.28          \\
\multirow{2}{*}{\textbf{1\&3\&4\&5\&6\&7}}            & \textbf{LSTM}          & \textbf{67.88} & \textbf{32.23} & \textbf{26.07} \\
                                                      & \textbf{BiLSTM}        & \textbf{68.61} & \textbf{33.97} & \textbf{26.78} \\
\multirow{2}{*}{2\&3\&4\&5\&6\&7}                     & LSTM                   & 67.78          & 32.11          & 25.87          \\
                                                      & BiLSTM                 & 68.43          & 33.74          & 26.58          \\ \hline
\end{tabular}
\caption{Comparison of Six Types of Feature}
\end{table}

\begin{table}[ht]
\centering
\begin{tabular}{lcccc}
\hline
\multicolumn{1}{c}{\multirow{2}{*}{Combinations of Seven Types of Features}} & \multirow{2}{*}{Model} & \multicolumn{3}{c}{Accuracy(\%)} \\ \cline{3-5} 
\multicolumn{1}{c}{}                                                         &                        & 2-class   & 5-class   & 7-class  \\ \hline
\multirow{2}{*}{All Seven Features}                                          & LSTM                   & 68.01     & 33.06     & 25.99    \\
                                                                             & BiLSTM                 & 68.67     & 34.18     & 26.12    \\ \hline
\end{tabular}
\caption{Comparison of Seven Types of Feature}
\end{table}

\end{document}